\documentclass[letter]{aa} 
\usepackage{graphicx}
\usepackage{txfonts}
\newcommand{\msol}{M$_\odot$}
\begin{document}
   \title{The distribution of stellar population age in galactic bars}

   \author{Herv\'e Wozniak
          \inst{1}
          }


   \institute{Universit\'e de Lyon, Lyon, F-69000, France ; Universit\'e Lyon 1,
Villeurbanne, F-69622, France ; Centre de Recherche Astronomique de
Lyon, Observatoire de Lyon, 9 avenue Charles Andr\'e, Saint-Genis Laval cedex, F-69561, 
France ;  CNRS, UMR 5574 ; \\ Ecole Normale Sup\'erieure de Lyon, Lyon, France\\
\\
              \email{herve.wozniak@obs.univ-lyon1.fr}
             }

   \date{Received December 26, 2006; accepted 22 January, 2007}

 
  \abstract
   {Recent analysis of stellar populations in barred galaxies have
    focused on the spatial distribution of stellar population ages and
    metallicities. However, barred galaxies are complex objects where
    dynamical instabilities play a leading role in shaping any spatial
    distribution.}
   {The age distribution of stellar populations should thus be
    analyzed from the two points of view of stellar population
    evolution and dynamical secular evolution.}
   {Chemodynamical simulations of single barred galaxies with
    simple but realistic star formation and feedback recipes are used to
    produce face-on mass-weighted maps of stellar population
    ages. Luminosity-weighted maps in V-band are also displayed after
    calibrating the simulation with mass-to-light ratios provided by a
    synthesis population model.}
   {It is shown that inside a stellar bar two persistent diametrically
   opposed regions display a mean age lower than the surrounding
   average. These two low-age regions are due to the accumulation of
   young stellar populations trapped on elliptical-like orbits along
   the bar, near the ultra-harmonic resonance. Age gradients along the bar 
   major-axis are comparable to recent observations. Another low-age region
   is the outer ring located near the Outer Lindblad Resonance, but
   the presence and strength of this structure is very time-dependent.}
   {}

   \maketitle
%

\section{Introduction}

Thanks to the improvement in stellar libraries, population synthesis
models, and inversion algorithms, dating of stellar populations in
galaxies and determination of star formation history has become a very
active field.  Indeed, the integrated spectra of galaxies provides a
huge number of information on the current stellar populations and the
history of their formation. Moreover, they simultaneously provide an
insight into their current dynamical state that results both from the
history of the mass assembling and their secular evolution.  However,
when the spectral information is combined with the spatial one, as for
integral field spectrographs or even long-slit spectroscopy of nearby
galaxies, one hits a snag because the spatial variation of the star
formation history cannot be understand without any knowledge on the
stars and gas dynamics.  It is thus difficult or even impossible to
unravel the two histories, the gravitational dynamics from the star
formation, since the physical mechanisms at play during any galaxy
life are so much intricately mixed. A direct approach, using N-body
and hydrodynamical simulations coupled with small scale recipes of
star formation and feedback, is thus useful to disentangle the two
histories even at the cost of very simplified physical assumptions.

In the zoo of galaxies, barred galaxies deserve a special attention
since they represent more than 75\%\ of all disc galaxies in the local
Universe. Most of them are still forming stars so that in the same
object we can find several generations of stellar populations, of
various ages and metallicities. The interpretation of age and/or
metallicity gradients in barred galaxies are thus not
straightforward. Using numerical simulations, Friedli et
al. (\cite{fbk94}) found that a bar is able to flatten abundance
gradients in the disc, in agreement with most observations of O/H
gradients. They did not investigate the effects of the bar on the age
distribution.

The main goal of this Letter is thus to provide qualitatively the age
distribution of stellar populations in typical barred galaxies.
Chemodynamical simulations, calibrated using Michel-Dansac \& Wozniak
(\cite{mw04}) technique, have been used to produce \emph{luminosity-}
and \emph{mass-}weighted maps of stellar population ages. I discuss
the origin and persistence of some features in the framework of the
orbital structure of stellar bars.


\section{A simple and generic chemodynamical model}

An {\em initial} stellar population is set up to reproduce a typical
disc galaxy. Positions and velocities for $2.5\,10^6$ particles are
drawn from a superposition of two axisymmetrical Miyamoto \& Nagai
(\cite{mn75}) discs of mass respectively $10^{10}$ and
$10^{11}$\msol, of scale lengths respectively $1$ and $3.5$~kpc and
common scale-height of $0.5$~kpc. Initial velocity dispersions are
computed solving numerically the Jeans equations. The initial disc
radius is 30~kpc.  The gaseous component is represented by 50\,000
particles for a total mass of $1.1\,10^{10}$~M$_{\sun}$ distributed in
a 6~kpc scale-length Miyamoto-Nagai disc.

Evolution is computed with a particle--mesh N-body code which includes
stars, gas and recipes to simulate star formation. The broad outlines
of the code are the following: the gravitational forces are computed
with a particle--mesh method using a 3D log--polar grid with $(N_R,
N_\phi, N_Z)=(60,64,312)$ active cells. The smallest radial cell in
the central region is 36~pc large and the vertical sampling is
50~pc. The extent of the mesh is 100~kpc in radius and $\pm 7.8$~kpc
in height.  The hydrodynamical equations are solved using the SPH
technique.  Since I used a polar grid, the pre-computation of
self-forces have been improved by subdividing each cell in $(n_r,
n_\phi, n_z)=(32,6,6)$ sub-cells. Self-forces are then linearly
interpolated before being subtracted from gravitational forces.  The
spatial and forces resolutions are thus much higher than in our
previous studies based on the same code (e.g. Michel-Dansac \&
Wozniak, \cite{mw04,mw06}).

The star formation process is based on Toomre's criterion for the
radial instability of gaseous discs (cf. Michel-Dansac \& Wozniak
\cite{mw04} for more details). When star formation is active, the
radiative cooling of the gas is computed assuming a solar metallicity.

This simulation has been chosen as being illustrative of a moderately
barred galaxy.

At the end of the simulation ($t\approx 3$~Gyr), the total number of
particles is roughly $3.2\,10^{6}$ for the stellar component and
$30\,000$ for the gaseous one. 46\%\ of the gas have been transformed
into stellar particles, mainly in the central 10~kpc. However, in this
relatively wide region, the new population (called 'young' hereafter)
accounts for roughly 5\%\ of the total mass (gas and stars) while in
the central kpc, the young population accounts for roughly 17\% of the
total mass.

This paper being focused on the expected age distribution, as inferred
from numerical simulations, a photometric calibration was performed as
in previous studies (e.g. Michel-Dansac \& Wozniak \cite{mw04,mw06})
using the information about the age and the metallicity of each
particle. I follow closely Michel-Dansac \& Wozniak (\cite{mw04}).
For each stellar particle, given its age and metallicity, the
mass-to-light ratio $\Upsilon$ is obtained from a bi-linear
interpolation into the tables of Bruzual \& Charlot (\cite{bc03}),
for a Salpeter initial mass function (IMF), with mass cut-off at 0.1
and 100 M$_{\sun}$.  Then, luminosity of the particle $i$ (${L}^{i}$),
in the V-band, is computed according to its mass (${M}^{i}$), birth
metallicity ($Z_{\mathrm{born}}^{i}$) and the time elapsed since its
birth ($t-t_{\mathrm{born}}^{i}$):
$${L}^{i}_{\mathrm{V}}(t) = \frac{{M}^{i}}
{\Upsilon_{\mathrm{V}}(t-t^{i}_{\mathrm{born}},Z^{i}_{\mathrm{born}})}.
$$ An age of 5~Gyr is assumed for the initial population at the
beginning of the simulation and a solar metallicity. Other assumptions
have been tried, in particular imposing an initial metallicity
gradient similar to our Galaxy or setting a younger or older initial
population, but I found that the results are qualitatively independent
of the choice.

\begin{figure}[t]
   \centering
   \resizebox{\hsize}{!}{\includegraphics{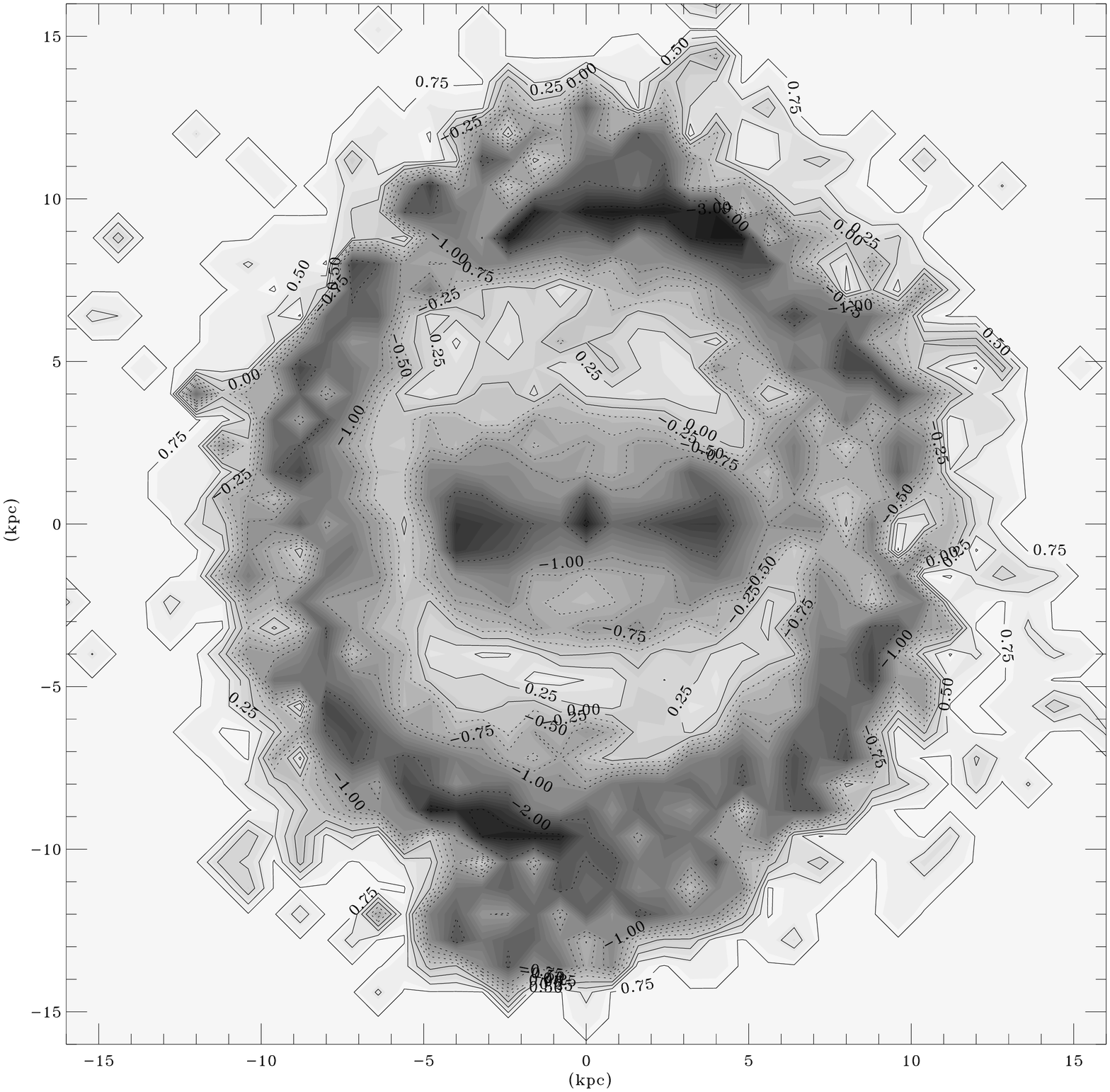}}
   \resizebox{\hsize}{!}{\includegraphics{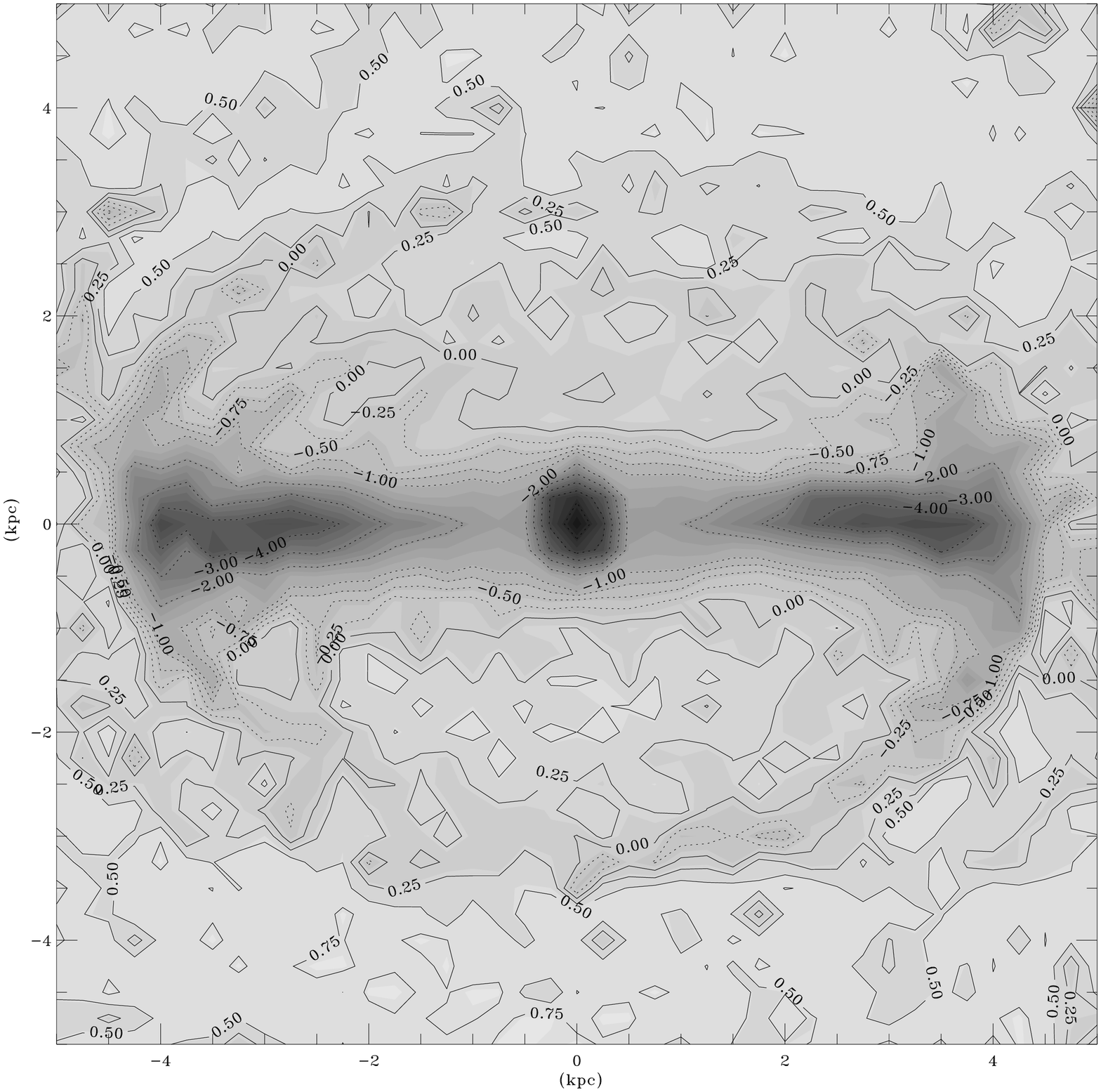}}
   \caption{Top panel: large scale V-\emph{luminosity}-weighted age. The
   field of view is 32~kpc wide. Regions younger than the mean are
   darker and isochronous curves are plotted as dotted lines. Regions
   older than the mean are brighter and isochrones are plotted as
   continuous lines. The number of standard deviation from the mean
   labels the isochrones, spaced by 0.25. Bottom panel: same as top
   panel but for a 10~kpc field of view.}
   \label{fig:lum_age}
   \end{figure}

\begin{figure}[t]
   \centering
   \resizebox{\hsize}{!}{\includegraphics{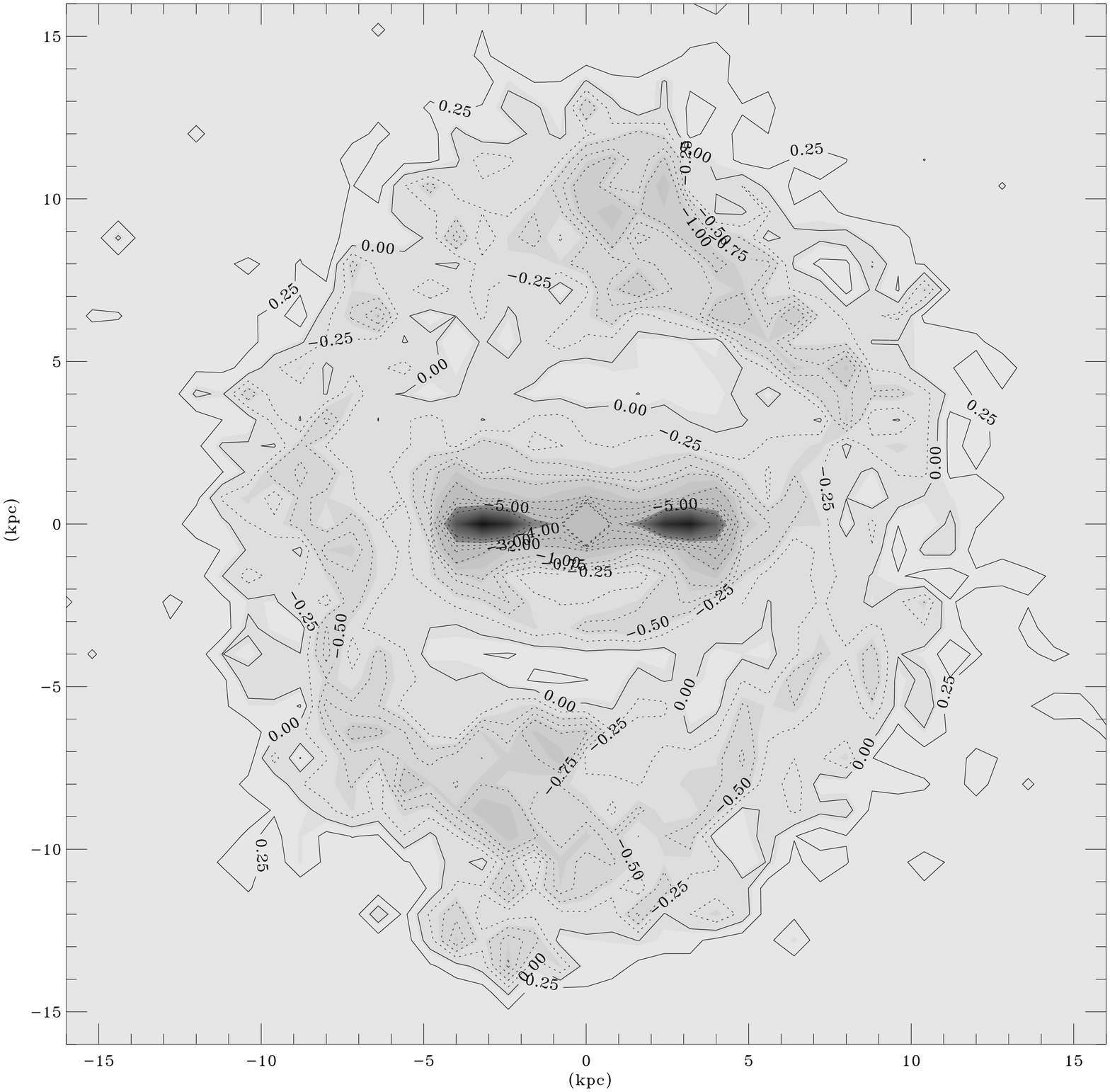}}
   \resizebox{\hsize}{!}{\includegraphics{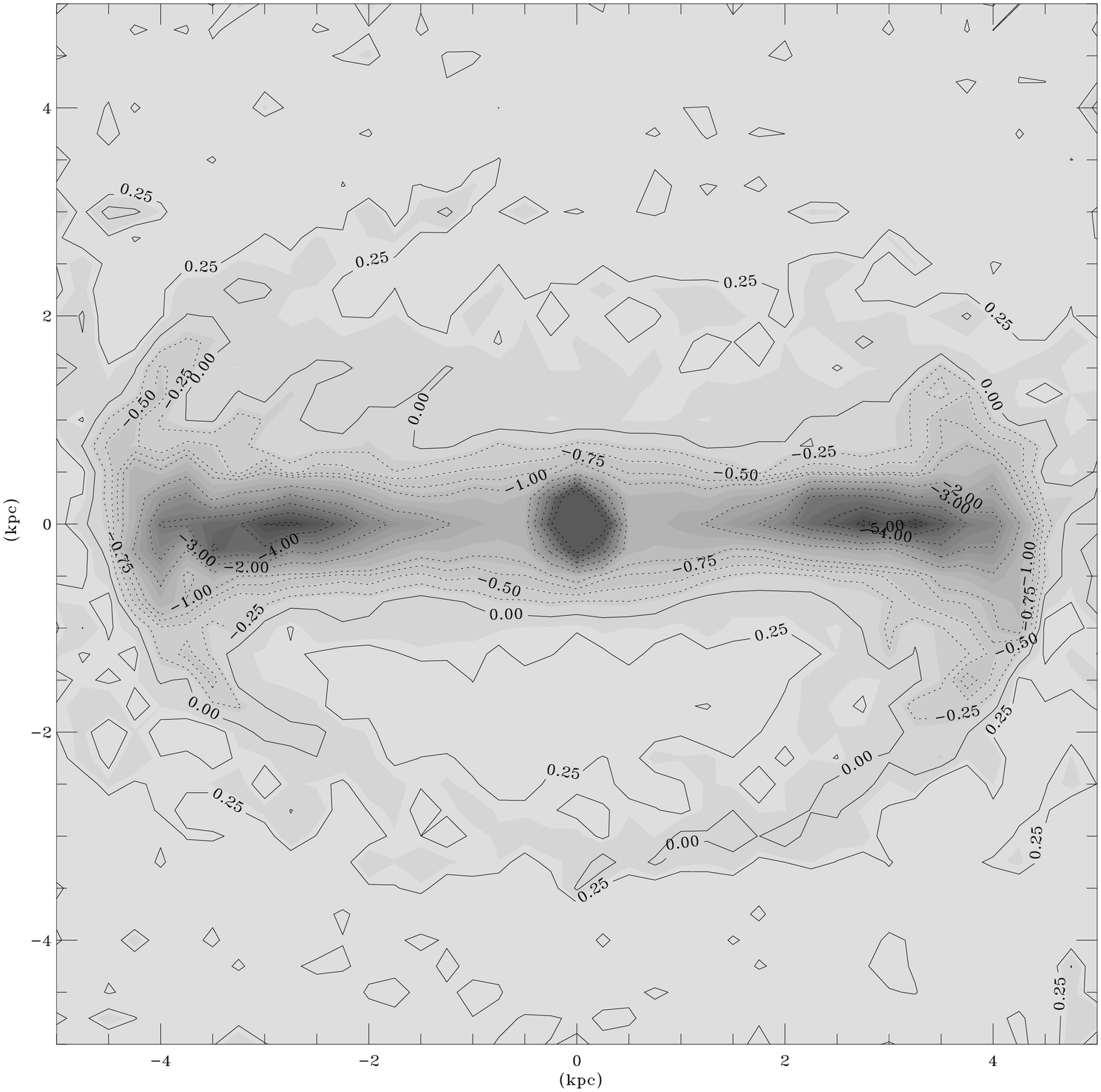}}
   \caption{Same as Fig.~\ref{fig:lum_age} but for
   \emph{mass}-weighted age.}
   \label{fig:mass_age}
   \end{figure}

Two-dimensional maps of the age distribution have been computed. A low
resolution of 41$\times$41 pixels has been chosen to smooth out the
many meaningless features due to the fluctuation of the number of
particles. For each pixel, the average age is the sum of each particle
ages weighted either by its mass or by its luminosity. These maps are
displayed in Fig.~\ref{fig:lum_age} and \ref{fig:mass_age} for the
particular time $T=1.2$~Gyr chosen to be illustrative of all others
snapshots.

%
\section{\emph{Mass}- and \emph{Luminosity}-weighted age distribution}

The ages should be use with caution. Indeed, most likely the initial
population should have a complex star formation history due in part to
the assembly of the galactic disk and bulge and in other part to its
internal dynamical secular evolution. However, to limit the number of
free parameters, all the initial population particles have the same
age. The amplitude of age differences from one place to another in the
simulation is thus meaningless since it obviously depends on the age
of the initial population. On the contrary, when a region is dominated
by the luminosity of a young population, it remains identified as a
low-age region whatever the age of the initial population because the
mass-to-light ratio of the young population is much lower than the old
one. For this reason, I plot in Fig.~\ref{fig:lum_age} and
\ref{fig:mass_age} the deviation from the mean age as a fraction of
the standard deviation. This highlights several regions of interest
whereas the basic qualitative results are thus independent of the
assumption made about the old (initial) population.

At large scale (Fig.~\ref{fig:lum_age} top), the two structures with
populations younger than the mean, are the stellar bar of $\approx
10$~kpc length and the outer ring (or pseudoring, see discussion in
the next Section) that encircles it. The youngest regions along the
ring are shifted by 10 to 20 degrees since star formation is provoked
by the compression in the enrolling gaseous spiral arms which are
trailing the bar. However, looking at various calibrated snapshots
between 0 and 3~Gyr, the distribution and morphology of the low-age
regions in the ring appear very time-dependent. Fig.~\ref{fig:lum_age}
is thus only a particular example of such a distribution.

At the scale of the bar (Fig.~\ref{fig:lum_age}), the young population
is mainly visible along the major-axis of the bar. Apart from the
central kpc, that is the site of a regular star formation due to the
gas inflow, the youngest regions are mainly located between roughly 2
and 4 kpc from the centre. On the contrary to the outer ring, these
low-age regions are persistent, at roughly the same place, that is
near the end of the stellar bar.  The morphology of these regions
slightly changes with time but not their location. At $T=1.2$~Gyr
(Fig.~\ref{fig:lum_age}), it is noteworthy that their outermost parts
looks like trailing spiral arms, the rotation being anti-clockwise.

It is noteworthy that the intermediate region between the bar and the
ring is depleted from a young and bright population. In this region
the age of the population is thus quite similar to that of the outer
disc, beyond the outer ring. The ages deviate by roughly $+0.5$
standard deviation from the map average.

The luminosity calibration in the V-band obviously enhances the region
of low-age particles since the mass-to-light ratio of these
populations is lower than for the initial one (assumed here to be
5~Gyr old at the beginning of the simulation, so 6.2~Gyr for the
snapshot displayed). The \emph{mass}-weighted is much more meaningful
for a dynamical analysis (cf. Fig.~\ref{fig:mass_age}).

%
\section{Discussion}

\begin{figure*}
   \centering \resizebox{\hsize}{!}{\includegraphics{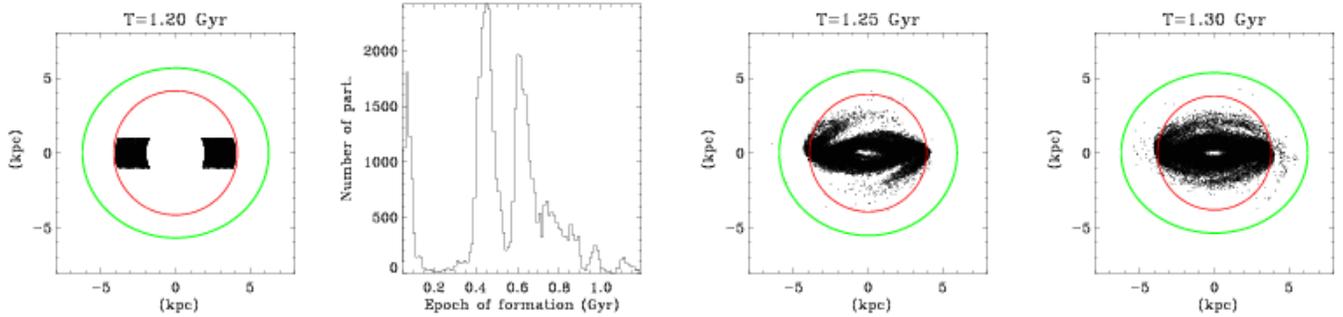}}
   \caption{Particles of the young population belonging to the regions
   along the stellar bar with the lowest mean age. From left to right:
   distribution of particles selected at $2 < R < 4$~kpc and $|y| <
   1$~kpc at $T=1.2$~Gyr; star formation history displayed as the
   distribution of the epochs of particle formation; particles
   positions after 50~Myr and 100~Myr. In all spatial plots, the bar
   has been rotated as to be aligned with the $x$-axis, the red circle
   is the ultra-harmonic radius, the green ellipse the corotation.}
   \label{fig:disperse}%
   \end{figure*}

It is noteworthy that the bar low-age regions are just inside or close
to the ultra-harmonic resonance (or 4/1, hereafter called UHR). In
Fig.~\ref{fig:disperse}, I select all the particles of the young
population for which $2 < R < 4$~kpc and $|y| < 1$~kpc at
$T=1.2$~Gyr. The linear UHR radius is $\approx 4.24$~kpc. The
formation history of these particles is obviously the staking of
several episodes of active star formation occurring at various places
in the galaxy as shown by the star formation history of these regions
(cf. Fig.~\ref{fig:disperse}). 

The fact that all these particles lie near the UHR along the
major-axis of the bar does not mean that they were born at this place
and will remain for a long time. Indeed, the bar dynamics is
particularly efficient in dissolving these regions since 50 and
100~Myr later, all these particles are spread out in a wider region.
Particles trapped by elliptical-like orbits (e.g. $x_1$ family in the
equatorial plane) remain inside a region approximately limited by the
UHR radius whereas particles with higher energy can fill a wider
region limited only by the corotation. For some snapshots, particles
have even enough energy to possibly cross over the corotation and lie
in the disc for a while (the so-called 'hot' population).

The explanation for the persistence of the low-age regions is then
straightforward. Indeed, particles trapped by elliptical-like families
of orbits spend most of their time at the apocentre of their
orbits. The accumulation of young particles on the inner side of the
UHR is thus simply due to the vanishing radial velocity in this
region. This is thus a pure effect of the bar dynamics.

   \begin{figure}
   \centering 
   \resizebox{7cm}{!}{\includegraphics{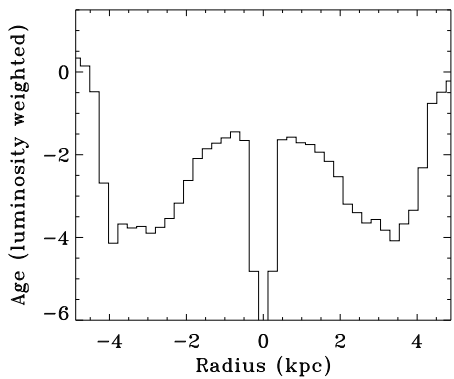}}
   \caption{ Cut of 360~pc wide along the bar major-axis through the
   V-\emph{luminosity}-weighted age distribution displayed in
   Fig.~\ref{fig:lum_age}. The width of the slit is similar to the
   observations of P\'erez et al. \cite{psz07}. See the text for the
   explanation of age units.}
   \label{fig:cut}%
   \end{figure}

These low-age regions have been already observed in a few barred
galaxies (P\'erez et al. \cite{psz07}; S. Courteau, L. MacArthur
private communication), using various age and metallicity indicators.
In particular, P\'erez et al. (\cite{psz07}) recently claimed that in
their sample of 6 barred galaxies, the outer parts of all stellar bars
are younger than the internal parts. To make a crude comparison with
their radial plots of H$\beta$ and H$\delta_A$ indices I display in
Fig.~\ref{fig:cut} a cut made through the age distribution map of
Fig.~\ref{fig:lum_age}. The width of the averaging `slit' has been
taken similar to observations, that is 360~pc wide. The age radial
profile displays a strong gradient as in some observations mentioned
above. However, a definitive comparison of population ages with
H$\beta$ or H$\delta_A$ age indicators certainly deserves an in depth
analysis using a spectral calibration of the simulation particles
which is beyond the scope of this Letter.  Moreover, a more detailed
comparison with these observations will certainly allow some tuning of
the initial population parameters of my chemodynamical simulations.

In the central kpc, the situation is quite more complex. Following the
previous reasoning, the tangential velocity of particles on $x_1$-like
orbits is maximum at the pericentre. The time they spend in the
central region is thus very short. However, several generations of
particles are created in the central kpc. This population is, by the
way, responsible of the so-called $\sigma$-drop phenomenon (Wozniak et
al.~\cite{wcef03}). But, trapped into the potential well, they mainly
remain confined in the circumnuclear region (Wozniak \& Champavert
\cite{wc06}). As long as the central region is forming stars, the mean
age will remain several standard deviation lower than the average.

Age gradients and spatial fluctuations are however not expected to be
very high except inside the bar as discussed above. Indeed, for
particles not trapped inside the UHR, the gravitational torques
exerted by the bar on the whole disc is a very efficient dynamical
process to disperse young population. Fig.~\ref{fig:disperse} shows
that the particles located in the low-age regions at $T=1.2$~Gyr are
rapidly scattered. The diffusion timescale is rather short, all
particles being mixed in less than 100~Myr. This is in agreement with
the findings of Friedli et al. (\cite{fbk94}) that the bar is
responsible to shallow any metallicity gradient outside corotation.

The low-age ring that encircles the bar must be undoubtedly associated
with the so-called \emph{outer} resonant ring phenomenon, since the
linear Outer Lindblad Resonance is located at roughly
8.84~kpc. According to Buta \& Combes (\cite{bc96}), this sort of rings
appear as blue components, sites of active star formation, on deep
images, in about one fifth of all spiral disk galaxies. It is
noteworthy that one third hosts pseudorings that are broken or partial
rings made up of spiral arms. The feature displayed in
Fig.~\ref{fig:lum_age} is thus rather a pseudoring than a ring.

\section{Conclusion}

Two diametrically opposed regions, located near the end of stellar
bars are expected to host a composite stellar population with a mean
age lower than the average. These two low-age regions have a dynamical
origin since they are due to the accumulation of young stellar
populations trapped in elliptical-like orbits along the bar, near the
ultra-harmonic resonance. 

The central region is also a low-age region because of the
sustained star formation activity generated by the gas accumulation
inside the ILR. Thus, between the nuclear region and the ends of a
bar, the population age displays strong gradients or varying slopes,
in agreement with recent observations.

Another low-age region is the outer ring located near the Outer
Lindblad Resonance, but the presence, strength and morphological
details of this structure is time-dependent.

\begin{acknowledgements}
I am very grateful to S. Courteau, L. MacArthur, P. S\'anchez-Bl\'azquez
and I. P\'erez for lively discussions on the age of stellar population
in barred galaxies during the IAU Symposium 241 in La
Palma. Computations have been performed on a 18 nodes cluster of PC
funded by the INSU/CNRS (ATIP No 2JE014 and Programme National
Galaxie).
\end{acknowledgements}


\begin{thebibliography}{}

\bibitem[2003]{bc03} Bruzual G., Charlot S., 2003, MNRAS, 344, 1000
\bibitem[1996]{bc96} Buta R., Combes F., 1996, \fcp\ 17, 95
\bibitem[1994]{fbk94} Friedli D., Benz W., Kennicutt R., 1994, ApJ 430, L105
\bibitem[2004]{mw04} Michel-Dansac L., Wozniak H., 2004, A\&A, 421, 863
\bibitem[2006]{mw06} Michel-Dansac L., Wozniak H., 2006, A\&A, 452, 97 
\bibitem[1975]{mn75} Miyamoto M., Nagai R., 1975, PASJ 27, 533
\bibitem[2006]{psz07} P\'erez I., S\'anchez-Bl\'azquez P., Zurita A.,
2006 A\&A submitted (astro-ph/0612159)
\bibitem[2006]{wc06} Wozniak H., Champavert N., 2006, MNRAS 369, 853
\bibitem[2003]{wcef03} Wozniak H., Combes F., Emsellem E., Friedli D.,
2003, A\&A, 409, 469

\end{thebibliography}
\end{document}